\documentclass[10pt]{iopart}
\usepackage{latexsym}
\usepackage{amscd}
\begin{document}
 \def\lambdabar{\protect\@lambdabar}
\def\@lambdabar{%
\relax
\bgroup
\def\@tempa{\hbox{\raise.73\ht0
\hbox to0pt{\kern.25\wd0\vrule width.5\wd0
height.1pt depth.1pt\hss}\box0}}%
\mathchoice{\setbox0\hbox{$\displaystyle\lambda$}\@tempa}%
{\setbox0\hbox{$\textstyle\lambda$}\@tempa}%
{\setbox0\hbox{$\scriptstyle\lambda$}\@tempa}%
{\setbox0\hbox{$\scriptscriptstyle\lambda$}\@tempa}%
\egroup}

\def\bbox#1{%
\relax\ifmmode
\mathchoice
{{\hbox{\boldmath$\displaystyle#1$}}}%
{{\hbox{\boldmath$\textstyle#1$}}}%
{{\hbox{\boldmath$\scriptstyle#1$}}}%
{{\hbox{\boldmath$\scriptscriptstyle#1$}}}%
\else
\mbox{#1}%
\fi
}
\newcommand{\beq}{\begin{equation}}
\newcommand{\eeq}{\end{equation}}
\newcommand{\beqan}{\begin{eqnarray*}}
\newcommand{\eeqan}{\end{eqnarray*}}
\newcommand{\beqa}{\begin{eqnarray}}
\newcommand{\eeqa}{\end{eqnarray}}
\newcommand{\hX}{\widehat{\bbox{X}}}
\newcommand{\hM}{\widehat{\bbox{M}}}
\newcommand{\hL}{\widehat{\bbox{L}}}
\newcommand{\hT}{\widehat{\bbox{T}}}
\newcommand{\hQ}{\widehat{\bbox{Q}}}
\newcommand{\eps}{\varepsilon}
\newcommand{\la}{\lambda}
\newcommand{\scr}{\scriptstyle}
\newcommand{\al}{\alpha}
\renewcommand{\d}{\partial}
\def\rmi{{\rm i}}
\def\rme{\bbox{\rm e}}
\def\rmd{\bbox{\rm d}}
\newcommand{\Bvec}{\bbox{B}}
\newcommand{\Evec}{\bbox{E}}
\newcommand{\Avec}{\bbox{A}}
\newcommand{\nablav}{\bbox{\nabla}}
\title{ Suppressing  and restoring  constants in physical equations}
\author{C. Vrejoiu }
\address{Facultatea de Fizic\v{a}, Universitatea din Bucure\c{s}ti, C.P.MG-11,
Bucure\c{s}ti-M\v{a}gurele, Rom\^{a}nia\\  E-mail : cvrejoiu@yahoo.com  }
\begin{abstract}
A simple procedure for restoring the constants in physical equations is introduced by 
a consistent consideration of the correspondence between the symbols representing 
physical quantities and pairs formed by numerical values and unit symbols. This procedure 
is applied to the very used unit systems stressing also the direct relations between the 
atomic,  natural systems and SI.  
\end{abstract}
\par To be submitted to Europ. J. Phys.
\vspace{1cm}
\par A frequent procedure for obtaining simple calculations consists in  setting some fundamental 
constants equal to unity. Sometimes, from physical interpretation reasons, it is 
necessary to restore these constants in the physical equations and in their solutions. 
This is just the point less treated in the textbooks (see for example \cite{bethe,land}).
 There are some general theoretical expositions in the literature on this 
problem \cite{desloge}
 but here we adopt a practical point of view trying to give some simple 
notation and prescriptions for restoring the constants in any expression. Illustrations 
for the very used unit systems as the atomic and natural ones are presented. 
Certainly, such notation and prescriptions are almost everyday discovered 
 and rediscovered by professors and 
students  teaching or learning physics. The present note,  not  claimed 
as an original one, is intended as a  didactic teaching aid for the beginners in physics. 
\par Let us an unit system labeled by $\al$. A physical quantity $\bbox{X}$ is specified in the 
given system by the correspondence $\bbox{X}\longrightarrow \{X_{\al},
\hX_{\al}\}$
where the first element of the pair is a number and the second one symbolizes the unit 
used for measuring $X$. By the same rule we can write $\hX_{\al}\longrightarrow 
\{X_{\al\beta},\hX_\beta\}$ relating the units of $\bbox{X}$ 
from two systems $\al$ and $\beta$ such that 
\beq\label{relunit}
X_\beta=X_\alpha X_{\al\beta}.
\eeq   
By writing the successive correspondences 
$$X\longrightarrow\left\{X_{\al},\hX_\alpha\right\}
\rightarrow\left\{X_{\al}X_{\alpha\beta},\hX_\beta\right\}
\rightarrow\left\{X_{\al}X_{\alpha\beta}X_{\beta\alpha},
\hX_\alpha\right\}$$
one obtains the simple and natural relation $X_{\beta\alpha}
X_{\al\beta}=1$.
\par Let a physical equation written in the $\al$-system as
$$F^{(\al)}\left(X^{(1)}_\alpha,\dots,X^{(n)}_\alpha\right)=0.$$
For passing to the $\beta$-system, we write
$$F^{(\al)}\left(X^{(1)}_\beta X^{(1)}_{\beta\alpha},\dots,X^{(n)}_\beta
X^{(n)}_{\beta\alpha}\right)=0$$
such that the rule for obtaining the equation written in the $\beta$-system is given 
by 
\beq\label{restor}
F^{(\al)}\left(X^{(1)}_\alpha,\dots,X^{(n)}_\alpha\right)=0
\stackrel{\scriptstyle{X_\alpha\rightarrow X_\beta X_{\beta\alpha}}}
{\longrightarrow}F^{(\beta)}
\left(X^{(1)}_\beta,\dots,X^{(n)}_\beta\right)=0.
\eeq
The following examples are the most encountered in the physics literature.
\par 1) {\it The atomic unit system} ($\al=a$) is introduced by the following values 
of electron's mass and charge and of the reduced Planck's constant: 
\beq\label{defua}
\left(m_e\right)_a=1,\;\left(e\right)_a=1,\;\left(\hbar\right)_a=1.
\eeq
Considering the relation between this system and the Gauss unit system
(cgsG: $ \al=G$), 
 these equations introduce the atomic units of mass, action and electric charge which, 
 expressed in cgsG units, are  represented by $M_{aG}=\left(m_e\right)_G,\,
 S_{aG}=\left(\hbar\right)_G,\,Q_{aG}=\left(e\right)_G.$
From the equations (\ref{defua}) we can write 
\beqan
\fl&~& \left(m_e\right)_{G}M_{Ga}=1,\;
 \left(e\right)^2_{G}M_{Ga}L^3_{Ga}
 T^{-2}_{Ga}=1 ,\;
 \left(\hbar\right)_{G}M_{Ga}L^2_{Ga}
 T^{-1}_{Ga}=1 
\eeqan
resulting the well known cgsG expressions of the atomic units as 

\beqa\label{G-ua}
\fl&~&M_{aG}=\frac{1}{M_{Ga}}=\left(m_e\right)_G,\,
L_{aG}=\frac{1}{L_{Ga}}=\left(\frac{\hbar^2}{m_ee^2}
\right)_{G},\,
T_{aG}=\frac{1}{T_{Ga}}=
\left(\frac{\hbar^3}{m_ee^4}\right)_{G}
\eeqa
$L_{aG}$ being the Bohr radius expressed in cgs units.
The  unit of the velocity $v$,   the value of  the vacuum light speed and the energy unit 
are given by  
\beqan
\fl&~&v_{aG}=\frac{1}{v_{Ga}}=\frac{L_{aG}}
{T_{aG}}=
\left(\frac{e^2}{\hbar}\right)_{G},\;
\left(c\right)_{a}=\left(c\right)_{G}v_{Ga}=\left(\frac{c\hbar}
{e^2}\right)_{G}=\frac{1}{\al}\approx 137\\
\fl&~&W_{aG}=\frac{1}{W_{Ga}}=M_{aG}L^2_{aG}T^{-2}_{aG}=
\left(\frac{m_ee^4}{\hbar^2}\right)_G\;\left(=2\left(Ry\right)_G\right).
\eeqan
 \par For passing to Heaviside-Lorentz unit system ($\al=H$), it suffices to give the 
 relation between the electric charge units $\hQ_H$ and $\hQ_G$: from Coulomb's law 
 written for the same charges and distances in the two unit systems we have 
 $F=Q^2_G/r^2=Q^2_H/4\pi r^2$ and, using equation (\ref{relunit}), $Q_G=Q_HQ_{HG}$, one obtains 
$ Q_{HG}=1/\sqrt{4\pi}$.  
   
\par We may pass from the atomic system to the SI one ($\al=I$) via the cgsG system by  well
known relations. Alternatively, one may define the atomic unit system starting from 
  SI by writing equations (\ref{defua}) with the SI numerical values for $m_e$, $e$, 
$\hbar$ and with the substitutions $G\longrightarrow I$. Because there are four basic 
units in SI 
\footnote{We consider only the mechanical and electromagnetic units} we can impose still another 
 condition. From Coulomb's law, $F=Q^2/4\pi\eps_0 r^2$, 
one sees that a reasonable condition is 
\beq\label{pieps}
\left(4\pi\eps_0\right)_a=1:\;\;4\pi\left(\eps_0\right)_IM^{-1}_{Ia}
L^{-3}_{Ia}T^2_{Ia}Q^2_{Ia}=1. 
\eeq
 The SI version of equations (\ref{defua}) furnishes the atomic units for mass and 
 electric charge
 \beq\label{mqau}
 M_{aI}=1/M_{Ia}=\left(m_e \right)_I,\;\;
 Q_{aI}=1/Q_{Ia}=\left(e \right)_I.
 \eeq
 Considering also equation (\ref{pieps}), one obtains the atomic units for length and 
 time:
 \beq\label{ltua}
\fl L_{aI}=\frac{1}{L_{Ia}}=\left(\frac{\hbar^2}{m_ee^2_0}\right)_I\;
 \left(=\left(Bohr's\; radius\right)_I\right),\;\;
 T_{aI}=\frac{1}{T_{Ia}}=\left(\frac{\hbar^3}{m_ee^4_0}\right)_I
 \eeq
 where $e_0=e/\sqrt{4\pi\eps_0}$.
 These results are obtained 
 from equations (\ref{G-ua}) by using the SI-numerical values for constants and by the 
 substitution $e\longrightarrow e_0$. 
\par It is easy to see that $\left(\mu_0\right)_a=\left(4\pi/c^2\right)_a$ and  Maxwell's equations 
in these atomic units are written as
\beq\label{Maxwua}
\nablav\times\Bvec=\frac{4\pi}{c^2}\bbox{j}+\frac{1}{c^2}\frac{\d\Evec}{\d t},\;
\nablav\times\Evec=-\frac{\d \Bvec}{\d t},\;\nablav\cdot\Bvec=0,\;\nablav\cdot\Evec=4\pi\rho.
\eeq   
 
 The same procedure is applicable 
 to the units of velocity and energy and for the vacuum light speed. For many applications 
 are useful the relations for the units of electric charge and current densities, electric and magnetic 
 fields and electromagnetic potentials:
 \beqa\label{EBua}
 \fl&~&\rho_{Ia}=\left(\frac{\hbar^6}{\sqrt{4\pi\eps_0}m^3_ee^7_0}\right)_I,\;
j_{Ia}=\left(\frac{\hbar^7}{\sqrt{4\pi\eps_0}m^3_e e^9_0}\right)_I,\;
E_{Ia}=\left(\frac{\sqrt{4\pi\eps_0}\hbar^4}{m^2_ee^5_0}\right)_I,\nonumber\\
\fl&~&  B_{Ia}=\left(\frac{e^2_0}{\hbar}\right)_IE_{Ia},\;\;
\Phi_{Ia}=\left(\frac{\sqrt{4\pi\eps_0}\hbar^2}{m_ee^3_0}\right)_I,\;
A_{Ia}=\left(\frac{\sqrt{4\pi\eps_0}\hbar}{m_ee_0}\right)_I
 \eeqa
 given as an exercise for the reader. It is also interesting to see how the equations (\ref{mqau}), 
 (\ref{ltua}) and (\ref{EBua}) are used for restoring the constants in Maxwell's equations 
 (\ref{Maxwua}) and obtaining the  corresponding SI equations. 
\par  Let a  simple example for the restoration of constants in the expression of the electromagnetic 
 energy density. The energy density associated with equations (\ref{Maxwua}) is 
 $\left(w\right)_a=\left(E^2+c^2B^2\right)_a/8\pi$. Applying equation (\ref{restor}),
 $$ \left(w\right)_I\frac{W_{Ia}}{L^3_{Ia}}=
 \frac{1}{8\pi}E^2_{Ia}\left[E^2+c^2B^2\right]_I.$$ Using the values given by equations (\ref{ltua}) and 
(\ref{EBua}), after simplifying the common factors we obtain the SI expression of the energy 
density $w=(\eps_0 E^2+B^2/\mu_0)/2$.

\par 2) {\it The natural system of units} ($\al=n$) is defined by the following values of 
 the reduced Planck's constant and of the vacuum light speed:
\beq\label{defnat}
\left(\hbar\right)_n=1,\;\;\left(c\right)_n=1.\eeq
These equations introduce directly the natural units of action and velocity expressed, for 
example, in cgsG units, as
$S_{nG}=\left(\hbar\right)_G,\,v_{nG}=\left(c\right)_G$.
Expressing the equations (\ref{defnat}) in cgsG system, one obtains
\beq\label{defnat-ec}
\fl \;\;\left(\hbar\right)_{G}M_{Gn}L^2_{Gn}
T^{-1}_{Gn}=1,\;\;
\left(c\right)_{G}L_{Gn}T^{-1}_{Gn}=1.
\eeq
Concerning the fundamental units of mass, length and time, one of these units may be 
chosen arbitrary and, usually, this is the length unit chosen as the Compton length 
of the electron: $L_{nG}=\left(\hbar/m_e c\right)_{G}.$
From these definitions one obtains\footnote{Equivalently we may chose the natural 
unit of mass as the electron mass.}
\beqa\label{cgs-un}
\fl&~&L_{nG}=\left(\hbar/m_ec\right)_{G},\;\;\;
 T_{nG}=\left(\hbar/m_ec^2 \right)_{G},
\;M_{nG}=\left(m_e\right)_{G}.
\eeqa
The natural unit of electric charge is obtained requiring the same formulation of 
Coulomb's law in cgsG and natural systems. We can write, using equation (\ref{restor}), 
$$F_G=Q^2_GR^{-2}_G\longrightarrow F_nF_{nG}=Q^2_nR^{-2}_nQ^2_{nG}L^{-2}_{nG}$$
and, because $F_n=Q^2_n/R^2_n$, 
\beq\label{un-q}
\fl Q_{nG}=\sqrt{\left(\hbar c\right)_{G}},\;\;
\left(e\right)_{n}=\left(e/\sqrt{\hbar c}\right)_{G}\approx 1/\sqrt{137},\;
\eeq
We can write also the following useful relations:
\beqa\label{EB-un-G}
 \fl&~& \rho_{nG}=\left(\sqrt{\hbar c}m^3_ec^3/ \hbar^3\right)_G,\;\;  
 j_{nG}=\left(c\right)_G\rho_{nG},\;\;
 E_{nG}=\left(\sqrt{\hbar c}m^2_ec^2/\hbar^2\right)_G,\nonumber\\
 \fl&~& B_{nG}=E_{nG},\;\;
 \Phi_{nG}= \left(\hbar/m_ec\right)_GE_{nG},\;
 A_{nG}=\Phi_{nG}.
 \eeqa

\par If equations (\ref{defnat}) are written in SI units one obtains the natural values 
$M_{nI},\;L_{nI},\;T_{nI}$ by the same procedure 
as in the case of cgsG:
$$
\fl L_{nI}=\left(\hbar/ m_ec\right)_{I},\;
T_{nI}=\left(\hbar /m_ec^2 \right)_{I},\;
M_{nI}=\left(m_e\right)_{I}.
$$
As in the atomic unit system case, we require $\left(4\pi\eps_0\right)_n=1$ resulting 
$\left(\mu_0\right)_n=4\pi$. Considering Coulomb's law, required to be expressed in 
natural units as $F=Q^2/r^2$,  
 one obtains $Q^2_{nI}=\left(4\pi\eps_0\hbar c\right)_I$ and 
for the electron charge 
$\left(e\right)_n=\left(e_0/\sqrt{\hbar c}\right)_I$ as in equation (\ref{un-q}).
 \par We can also derive the SI values of the units for the electric charge and 
 current densities, electric and magnetic fields and electromagnetic potentials:
 \beqa\label{EA-u-I}
 \fl&~& \rho_{nI}=\left(\sqrt{4\pi\eps_0}\sqrt{\hbar c}m^3_ec^3/
 \hbar^3\right)_I,\;\;  j_{nI}=\left(c\right)_I\rho_{nI},\;\;
 E_{nI}=\left(\sqrt{\hbar c}m^2_ec^2/\sqrt{4\pi\eps_0}
 \hbar^2\right)_I,\nonumber\\
 \fl&~& B_{nI}=\left(1/c\right)_IE_{nI},\;
 \Phi_{nI}= \left(\hbar/m_ec\right)_IE_{nI},\;
 A_{nI}=\left(1/c\right)_I\Phi_{nI}
 \eeqa
Let us, as an example, the Dirac equation for a charged particle in an external 
electromagnetic field  
written in natural units $\left(\rmi\hat{\d}-q\hat{A}-m\right)\psi(x)=0,\;
\left(A^\mu\right)=\left(\Phi,\,\Avec\right)$. Passing to the cgsG or SI units, $\al=G$ 
or $\al=I$, we write
\beqan
\fl&~& \left[\left(\rmi\hat{\d}-q\hat{A}-m\right)\psi\right]_n=0\longrightarrow\\ 
\fl&~&\left\{\frac{\rmi}{L_{\al n}}\hat{\d}-Q_{\al n}\,q
\left[\gamma^0\Phi\Phi_{\al n}-\bbox{\gamma}\cdot\Avec
A_{\al n}\right]-mM_{\al n}\right\}_\alpha\left(\psi\right)_\alpha=0.
\eeqan
 For $\al=G$, using equations (\ref{EB-un-G}) one obtains the Dirac equation in cgsG units 
$\left(\rmi\hbar\hat{\d}-(q/c)\hat{A}-mc\right)\psi=0,\;\left(A^\mu\right)=\left(\Phi,\,
\Avec\right)$ and, for $\al=I$, the same equation in SI units 
$\left(\rmi\hbar\hat{\d}-q\hat{A}-mc\right)\psi=0,\;\left(A^\mu\right)=\left(\Phi/c,\,
\Avec\right)$.
\vspace{1cm}
\par 

{\bf References}
\begin{thebibliography}{10}
\bibitem{bethe}
Bethe H, Salpeter E 1977  {\it Quantum Mechanics of One and Two Electron Atoms} 
Plenum Publ. Corp.,New York

\bibitem{land}
Landau L, Lifchitz E 1977  {\it Quantum Mechanics (nonrelativistic theory)} Pergamon Press, 
Oxford 
\bibitem{desloge}
Desloge E  1984 {\it Suppression and restoration of constants in physical 
equations} Am. J. Phys. {\bf 52}, 312
\end{thebibliography}
\end{document}